# Unveiling the Electric-current-limiting and Photodetection Effect in Two-dimensional Hydrogenated Borophene

Yipeng An,[1,2,*] Yusheng Hou,[2] Hui Wang,[2,3] Jie Li,[2] Ruqian Wu,[2,*] Tianxing Wang,[1] Haixia Da,[4] and Jutao Jiao[1]

[1]*College of Physics and Materials Science & International United Henan Key Laboratory of Boron Chemistry and Advanced Energy Materials, Henan Normal University, Xinxiang, Henan 453007, China*

[2]*Department of Physics and Astronomy, University of California, Irvine, California 92697, USA*

[3]*Hunan Key Laboratory of Super Microstructure and Ultrafast Process, School of Physics and Electronic, Central South University, Changsha, Hunan 410083, China*

[4]*College of Electronic and Optical Engineering & College of Microelectronics, Nanjing University of Posts and Telecommunications, Nanjing, Jiangsu 210046, China*

**ABSTRACT:**

The electronic transport and photoelectric properties of hydrogenated borophene $B_4H_4$, which was realized in a recent experiment by Nishino, et al. [J. Am. Chem. Soc. **139**, 13761 (2017)], are systematically investigated using the density functional theory and non-equilibrium Green's function methods. We find that $B_4H_4$ exhibits a perfect current-limiting effect and has high (along the zigzag direction) and low (along the armchair one) optional levels due to its strong electrical anisotropy. Moreover, $B_4H_4$ can generate sizable photocurrents under illumination, with strong photoelectronic response to blue/green light along the zigzag/armchair direction. Our work demonstrates that $B_4H_4$ is promising for the applications of current limiter and photodetectors.

---

* ypan@htu.edu.cn; wur@uci.edu

# I. INTRODUCTION

Recently, borophene and its derivatives have received increasing attention since its successful realization in 2015 [1,2]. Many different types of borophene-based monolayer (ML) [3-6] structures have been proposed in theory, and they are promising for a variety of applications [7], including Dirac materials [8], lithium-sulfur batteries [9], spin-filters [10], superconducting materials [11], hydrogen storage media [12], and catalysts [13]. The borophene MLs such as the out-of-plane buckling sheets [1] and $\beta_{12}$ and $\chi_3$ boron sheets [2] are often fabricated on metal (Ag) substrates, as a truly free-standing two-dimensional (2D) borophene is unstable according to theoretical studies [14,15]. Many fabrication strategies have been proposed to produce free-standing 2D borophene, such as using transition metal adatoms [16-18] and hydrogenation to saturate the dangling bonds [3,14,15,19].

Several stable hydrogenated borophene structures, such as the *C2/m*, *Pbcm*, and *Cmmm* phases, were recently predicted [3,14,20,21]. In particular, the *Cmmm-phase* $B_4H_4$ ML is a Dirac ring material [20] with a mechanical anisotropy along the zigzag and armchair directions, and has a high thermal conductivity [21]. The *Cmmm*-phase $B_4H_4$ ML has been successfully fabricated with exfoliation and complete ion-exchange method at room temperature [22]. To explore the potential of utilizing this material in nanodevices, further investigations of 2D $B_4H_4$ MLs are needed, especially for its electronic transport and photoelectronic response properties that have been largely left unveiled. Several important aspects that need to be thoroughly examined include: (a) does the 2D $B_4H_4$ ML have any peculiar current−voltage (*I−V*) behavior? (b) how strong is its electrical anisotropy? (c) what is its photoelectronic response under illumination? (d) is there any unique feature for potential device application?

In this paper, we systematically study the electronic transport and photoelectronic response properties of a 2D $B_4H_4$ ML (see Figure 1(a)) by using the first-principles approach. We find that the 2D $B_4H_4$ ML exhibits a strong electrical anisotropy along the zigzag and armchair directions. It exhibits a significant current-limiting (CL) effect and has two optional levels, i.e., high-level along the zigzag direction (z-$B_4H_4$) and low-level along the armchair one (a-$B_4H_4$). Moreover, $B_4H_4$ can generate sizeable photocurrents in blue (green) light along the zigzag (armchair) direction.

## II. METHODS

The electronic structures, electronic transport, and photoelectronic response properties of the 2D $B_4H_4$ ML are determined by using the density functional theory and nonequilibrium Green's function approach, as implemented in the Atomistix Toolkit (ATK) [23-25]. The exchange and correlation effect among electrons is described with the generalized gradient approximation (GGA) [26,27]. The B-1s core electrons are represented by the optimized Norm-Conserving Vanderbilt (ONCV) pseudo-potentials. The wave functions of valence states are expanded as linear combinations of atomic orbitals (LCAO), at the level of SG15 [28] pseudo-potentials and basis sets, which are fully relativistic and can provide comparable results as the all-electron approach. We use an energy cutoff 200 Ry for the basis expansion. The positions of all atoms in the unit cell are relaxed until the residual force on each atom is less than 0.005 eV/Å and the total free-energy tolerance is below $10^{-6}$ eV, respectively. Then the two-probe structures are constructed by repeating the unit cell. For the electronic transport calculations, we use a 1×7×150 Monkhorst-Pack *k*-points grid to sample the 2D Brillouin zone of the electrodes, to achieve a balance between the cost and accuracy.

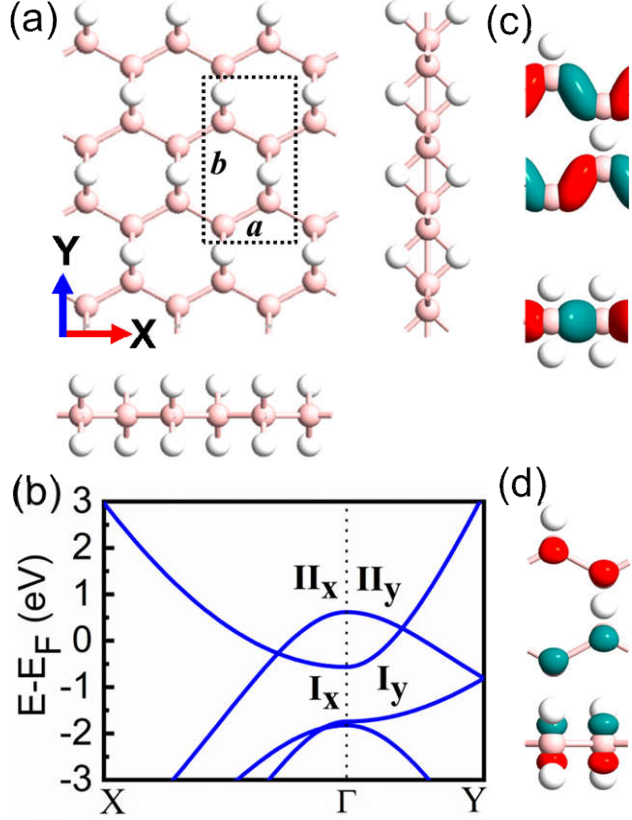

FIG. 1. (a) Structure of 2D $B_4H_4$ monolayer (including top and side views). (b) Band structures of rectangular $B_4H_4$ along the $\Gamma$–$X$ and $\Gamma$–$Y$ directions. Top and side views of eigenstates of bands **I** (c) and **II** (d) at $\Gamma$ point. The isovalue is set to 0.25 Å$^{(-3/2)}$.

## III. RESULTS AND DISCUSSION

Figure 1(a) shows the top and side views of 2D $B_4H_4$ ML. After geometry optimization, the lattice parameters *a* and *b* of the rectangular $B_4H_4$ unit cell are 3.000 and 5.299 Å, respectively, both in good agreement with the previous report [20]. It is easier to investigate its electrical anisotropy using a rectangular unit cell, including constructing a two-probe structure and performing transport calculations [29,30]. Figure 1(b) shows the band structures along the zigzag ($\Gamma$–$X$) and armchair ($\Gamma$–$Y$) directions, respectively. There are two bands (i.e., **I** and **II**) crossing the Fermi level ($E_F$) along these two directions. The hole effective mass of band **II** shows an obvious anisotropy, which can lead to an anisotropic current of a monolayer material (e.g., phosphorene) [31]. It is smaller along the zigzag direction ($m_x = 0.32m_0$) than along the armchair one ($m_y = 0.56m_0$). As a result, one may expect larger electron transmission along the zigzag direction. In addition, the eigenstates of bands **I** and **II**, which are composed of the hybrid $p_x$ orbitals of B atoms, are widely distributed on the zigzag B chains (see Fig. 1(c)) and very beneficial to the electron transmission. Except that, the eigenstates of band **II** at the $\varGamma$ point are composed of the $p_z$ orbitals of B atoms and very localized (see Fig. 1(d)), which are unbeneficial to the electron transmission? Test calculations indicate that inclusion of van der Waals correction doesn't change the electronic and transport properties of $B_4H_4$ ML.

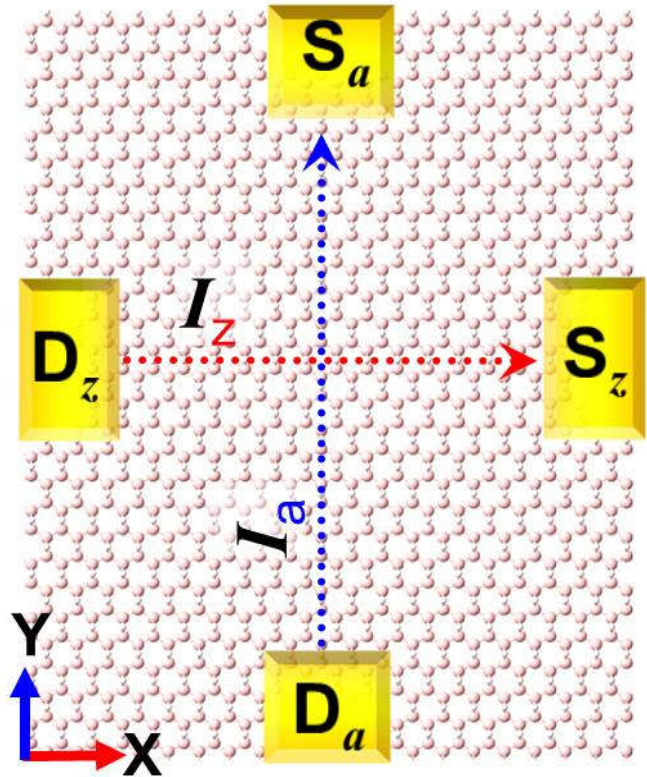

FIG. 2. Schematic of 2D $B_4H_4$ diode nanodevice. $D_{z/a}$ and $S_{z/a}$ refer to the drain and source electrodes along the zigzag (X axis)/armchair (Y axis) direction. $I_{z/a}$ shows the current along the zigzag/armchair direction.

To investigate the electrical transport properties, we construct the two-probe structure of 2D $B_4H_4$ (see Fig. 2) and directly calculate its electronic conductivities along the zigzag (X axis) and armchair (Y axis) directions. This 2D structure has a periodicity perpendicular to the transport direction between the drain (D) and source (S) electrodes. The third direction is out of the plane, along which the slabs are separated by a vacuum (more than 15 Å). Both the drain and source electrodes are described by a large supercell and are semi-infinite in length along the transport direction. We use the Dirichlet boundary conditions along the transport direction and periodic boundary conditions in the other two orthogonal directions. When a drain-source bias $V_b$ is applied across the D and S electrodes, their Fermi energies are shifted accordingly. A positive bias gives rise to an electric current from the D electrode to the S electrode, and vice versa. In the present work, the current $I$ through the z-$B_4H_4$ and a-$B_4H_4$ diode structures is determined by using the Landauer–Büttiker formula [32]

$$I(V_b) = \frac{2e}{h} \int_{-\infty}^{\infty} T(E, V_b)[f_D(E - \mu_D) - f_S(E - \mu_S)]dE, \qquad (1)$$

where $T(E, V_b)$ is the bias-dependent transmission coefficient, calculated from the Green's functions; $f_{D/S}$ are the Fermi-Dirac distribution functions of the drain/source electrodes; $\mu_D$ (= $E_F - eV_b/2$) and $\mu_S$ (= $E_F + eV_b/2$) are the electrochemical potentials of the drain and source electrodes respectively, and [$\mu_D$, $\mu_S$] defines the bias window (BW). More details of this method can be seen in the previous literatures [23-25].

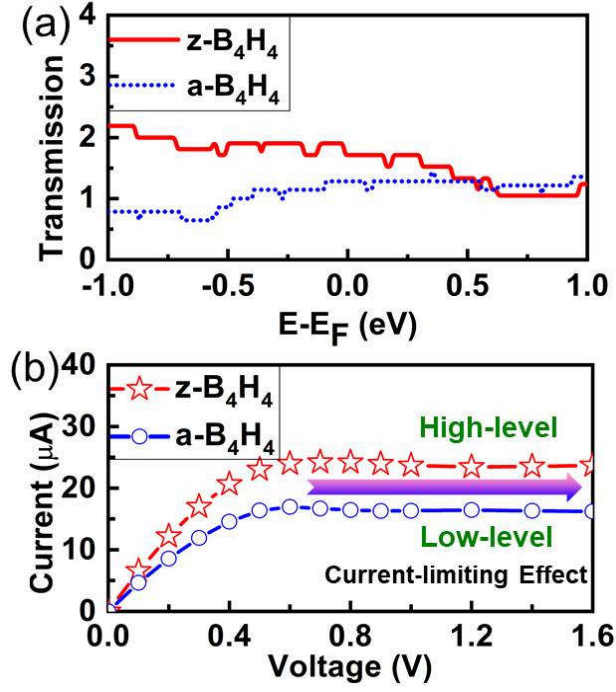

FIG. 3. (a) Transmission spectra at zero bias and (b) $I-V$ curves of z-$B_4H_4$ and a-$B_4H_4$. The Fermi level is shifted to zero.

Figure 3(a) shows the transmission spectra of z-$B_4H_4$ and a-$B_4H_4$ at zero bias, both of which have several steps and show an obvious quantized characteristic, analogous to graphene and other graphene-like structures [33-35]. It appears to be easier for electrons within a broad energy region (below 0.43 eV) to transmit along the zigzag direction due to their larger transmission coefficients, except in the high positive energy region (above 0.43 eV). As a result, z-$B_4H_4$ has larger equilibrium conductance (about 1.7 $G_0$) than a-$B_4H_4$ (about 1.3 $G_0$). Here, we only show results of z-$B_4H_4$/a-$B_4H_4$ with 14/10 rectangular unit cells in the scattering region. Nevertheless, our test calculations indicate that further length expansion gives little influence on the transport properties and we will focus our following discussions with results from the 14/10 unit cells.

Figure 3(b) displays the $I-V$ curves of z-$B_4H_4$ and a-$B_4H_4$ under biases from 0 to 1.6 V. Interestingly, they both show a perfect current-limiting effect when the threshold voltage is beyond 0.6 V. This has a useful application in electrical circuits by imposing an upper limit on the current for the protection purpose. This effect is characterized by two key factors that depend on materials and may vary substantially in experiments. The first factor is the threshold voltage where the current-limiting effect appears. In general, it is desired to have low CL voltage for minimizing the power consumption. The second one is the value of saturation current, which mainly determines the performance index of a CL device. The $B_4H_4$-based CL nanodevice has a threshold bias of 0.6 V and large saturation current with two optional levels, i.e., high-level (24.0 µA) along the zigzag direction and low-level (16.5 µA) along the armchair one. Thus, the $B_4H_4$ shows a strong electrical anisotropy, and its ratio of current anisotropy $\eta=I_z/I_a$ ($I_z$ and $I_a$ refer to the saturation current of z-$B_4H_4$ and a-$B_4H_4$, respectively) is about 1.5, equal to that of χ-borophene [36]. Note that the conductance of $B_4H_4$ decreases as the bias increases, and its maximum conductivity is about $1.5\times10^5$ S/m, less than that of Al ($3.8\times10^7$ S/m), Au ($4.5\times10^7$ S/m), and Cu ($5.9\times10^7$ S/m). Therefore, the $B_4H_4$ ML is a good candidate for the use in current limiters.

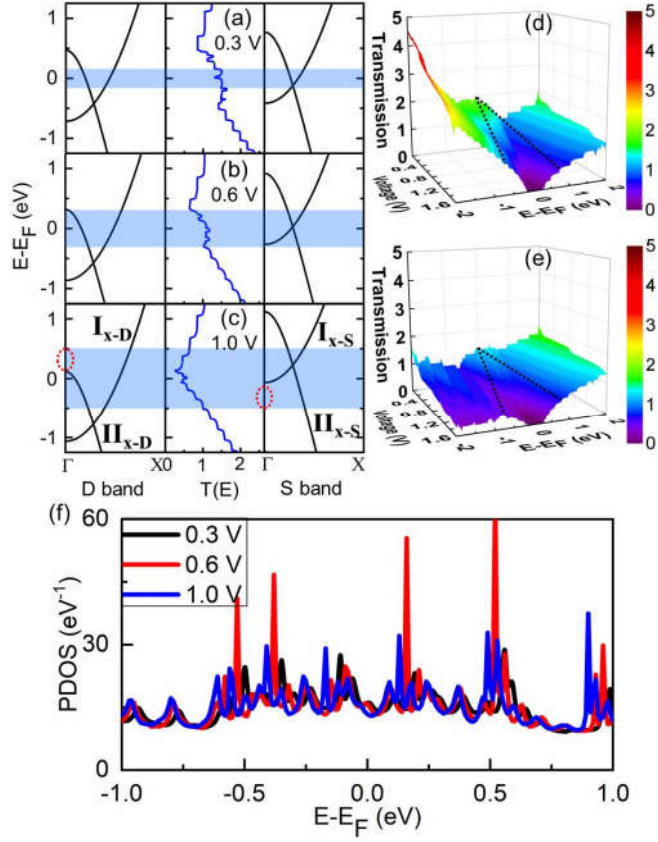

FIG. 4. (a)-(c) show the transmission spectra and band structures of drain and source electrodes of z-$B_4H_4$ at 0.3 (a), 0.6 (b), and 1.0 V (c), respectively. Transmission spectra of z-$B_4H_4$ (d) and a-$B_4H_4$ (e) at various biases. (f) PDOS of z-$B_4H_4$ at 0.3, 0.6, and 1.0 V, respectively. The shadows in (a)-(c) and black dashed lines in (d) and (e) are the bias windows. The red ellipses in (c) indicate the entrance of band into the bias window.

To unveil the physical origin of the current-limiting effect of $B_4H_4$, we analyze its band structures and transmission spectra under varying biases. In general, the electronic transport of monolayers is mainly determined by their band structures from the intra- and inter-band transitions around the Fermi level [37]. One side, only the electrons in the states between the bias window can contribute to the current [33,37]. For another, the band properties, including the parity limitation [33], the projection of element [38], and its localization, also play a key role in the electronic transport. Figures 4(a)-(c) show results of the z-B$_4$H$_4$ under biases of 0.3, 0.6, and 1.0 V, respectively. As a forward bias is applied to the drain and source electrodes, their bands shift down and up accordingly. The electron transmission of B$_4$H$_4$ ML is mostly dependent on the band overlaps between the bands **I** and **II** of the drain and source electrodes, as well as its band localization. As the bias increases, such as from 0 to 0.6 V, the band overlaps are gradually increasing. (see Figs. 4(a) and 4(b)). The electron transmission probability is similar (see Fig. 4(d)) due to their consistent band properties. Thus, its integral over the BW according to Eq. (1) (i.e., current) increases monotonically as the bias windows expands (see Fig. 3(b)). Note that the transmission coefficients near the energies of **II**$_{x-D}$ at the $\Gamma$ point are always smaller due to its localized eigenstates at $\Gamma$ (see Fig. 1(d)). When the bias is beyond 0.6 V, the band overlaps change little. But the localized eigenstates (at $\Gamma$) of band **II**$_{x-D}$ enter into the bias window (see Fig. 4(c)), leading to the obvious decrease of transmission coefficients. That is, the transmission around the Fermi level drops as a gap develops in the scattering region under the influence of a high electric field. This is shown by the evolution of the projected density of states (PDOS) of z-B$_4$H$_4$ in the scattering region as the bias increases from 0.3 to 0.6 and 1.0 V (see Fig. 4(f)). These give rise to the saturation of current beyond 0.6 V (i.e., current-limiting effect) of z-B$_4$H$_4$ even though the BW expands. The same CL mechanism is applicable to the a-B$_4$H$_4$. As the band **II**$_x$ of z-B$_4$H$_4$ and band **II**$_y$ of a-B$_4$H$_4$ have different effective masses, a-B$_4$H$_4$ has a less efficient electron transmission (see Fig. 4(e)) and lower $I-V$ curve (see Fig. 3(b)). Then the electrical anisotropy appears, and the high-level/low-level current-limiting effect is achievable along the zigzag/armchair direction.

The electric conduction can be resolved into contributions through different pathways in the system [39]. The local current between all pairs of atoms A and B, where A is on one side of the two-probe structure and B is on the other, equals the total current:

$$I(V_\text{b})= \sum_{i\in A, j\in B} I_{ij}(V_\text{b}) \qquad (2)$$

The total transmission coefficient can also be resolved to local bond contributions, $T_{ij}$, across the boundary between two parts (A and B atoms) as

$$T(E)= \sum_{i\in A, j\in B} T_{ij}(E) . \qquad (3)$$

To further understand the physical picture of electronic transport of $B_4H_4$ at the atomic level, we analyze its electron transmission pathways and corresponding atomic orbital origin. Generally, there are two types of local current pathways: *via* electron hopping (i.e., hopping current) between atoms or *via* chemical bonds (i.e., bond current) [40]. Our results show that B-B bonds play the dominant role for both z-$B_4H_4$ and a-$B_4H_4$. The bond current of z-$B_4H_4$ is along the zigzag chains parallel to the transport direction (see Fig. 5(a)). However, the bond current of a-$B_4H_4$ is mainly along the "step-like" pathways, which is unparallel to the transport direction but with an angle (see Fig. 5(b)). It reduces the electron transmission probability and suppresses the $I-V$ curve along the armchair direction.

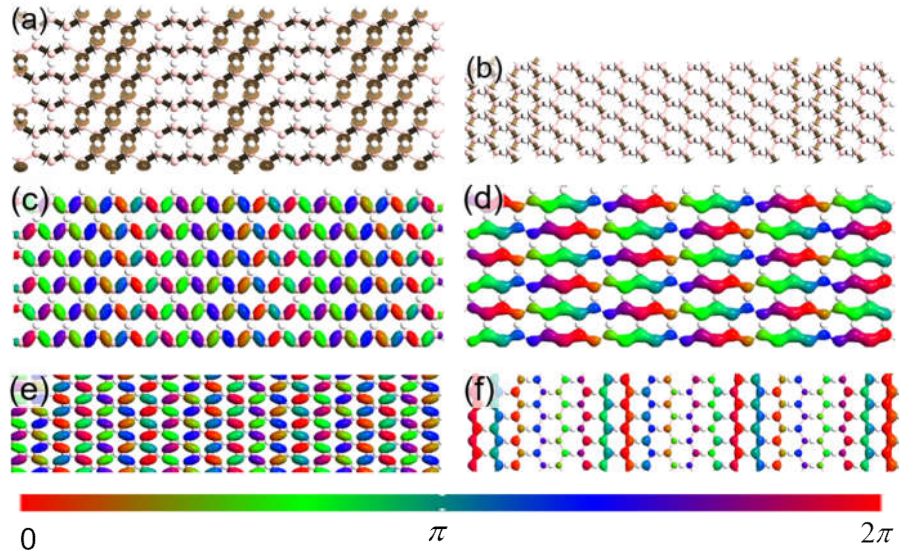

FIG. 5. Transmission pathways of z-$B_4H_4$ (a) and a-$B_4H_4$ (b) at the $E_F$ under the bias of 0.6 V. The arrows point in the directions of electric current flow. Corresponding transmission

eigenstates, TE-I (c) and TE-II (d) of z-$B_4H_4$, TE-I (e) and TE-II (f) of a-$B_4H_4$.

The transmission coefficient of z-$B_4H_4$ at the $E_F$ under the bias of 0.6 V is about 1.2 (see Fig. 4(b)), implying that there are two degenerate transmission channels (because this is a single-electron transmission process with a transmission probability no more than 1 for each channel). Figures 5(c) and 5(d) give the eigenstates of the two transmission channels (i.e., TE-I and TE-II) of z-$B_4H_4$, both are very delocalized and spread over the B atoms along the zigzag $B_4H_4$ chains and contribute to the electron transmission. For the a-$B_4H_4$, it also has two degenerate transmission eigenstates despite its transmission coefficient is about 0.8, less than z-$B_4H_4$. However, the TE-II of a-$B_4H_4$ (see Fig. 5(f)) is quite localized, only its TE-I (see Fig. 5(e)) has effective contribution to the current, which is unbeneficial to its electron transmission and causes its depressed $I-V$ curve (see Fig. 3(b)).

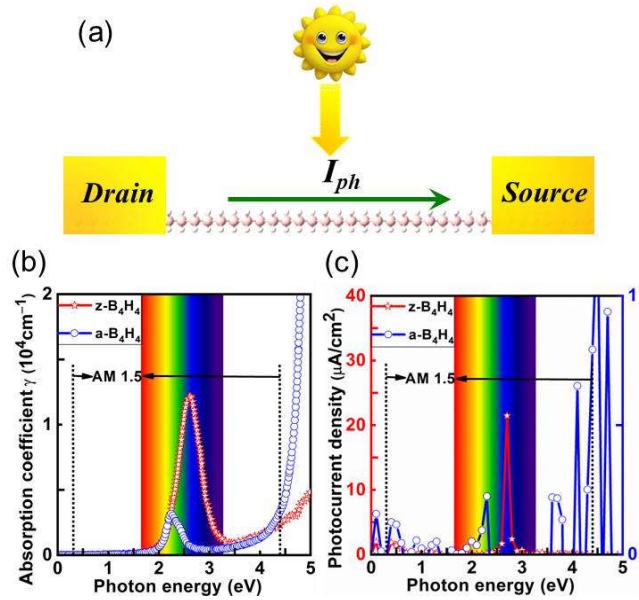

FIG. 6. (a) Schematic of $B_4H_4$ photodetector. (b) Absorption coefficient $\gamma$ and (c) photocurrent density of the z-$B_4H_4$ and a-$B_4H_4$ photoelectronic nanodevices. The embedded spectrum pattern shows the visible light region.

The monolayer nanomaterials may also find potential application in the fast-growing photoelectronics. For instance, the graphene-based photodetectors had been fabricated in experiments [41-45]. The large electrical anisotropy of $B_4H_4$ ML inspires us to further explore its photoelectronic properties. Figure 6(b) shows the photon absorption coefficient $\gamma$ as a function of photon energy along the zigzag and armchair directions. Within the visible light region and the AM1.5 standard [46], both the z-$B_4H_4$ and a-$B_4H_4$ have a strong absorption peak but appeared at different photon energy positions. Namely, the z-$B_4H_4$ is favorable to absorb the blue light (near 2.62 eV photon energy) with a large coefficient $\gamma=1.2\times10^4$ cm$^{-1}$, while a-$B_4H_4$ prefers to absorb the green one (near 2.24 eV) with a coefficient $\gamma=0.3\times10^4$ cm$^{-1}$. Note that in the ultraviolet region, $B_4H_4$ (especial along the armchair direction) has stronger photon absorption. It is hence indicated that the $B_4H_4$ could can be used in the photoelectronic devices, such as photodetectors.

Several 2D materials have been predicted as candidates of photoelectronic devices [30,47,48]. In particular, a few 2D-material-based (e.g., phosphorene and palladium diselenide) photodetectors have been predicted in recent reports [48,49]. Black phosphorus mid-infrared photodetectors and graphene heterojunction infrared photodetectors have already been realized in experiments [45,50,51]. To further explore the potential application of $B_4H_4$ in photoelectronics, we construct a $B_4H_4$-based photodiode (see Fig. 6(a)), and calculate its photocurrent along the zigzag and armchair directions by means of a theoretical method that was developed based on density functional theory within the nonequilibrium Green's function approach [29,52,53]. Using this method, we can obtain the photocurrent that flows through a photodiode composed of the drain and source electrodes and the central scattering region when it is irradiated by polarized light. More specifically, the photocurrent can be calculated as a first-order perturbation to the electronic system, originating from the interaction with a weak electromagnetic field. The electron-photon interaction is described by the Hamiltonian

$$H' = \frac{e}{m_0} \mathbf{A} \cdot \mathbf{P}, \tag{4}$$

where **A** is the vector potential and **P** is the momentum operator. The photoexcited current into electrode α = D/S due to absorption of *N* photons with frequency *ω* is obtained by

$$I_\alpha = \frac{e}{h} \int_{-\infty}^{\infty} \sum_{\beta=D,S} [1 - f_\alpha(E)] f_\beta(E - \omega) T_{\alpha,\beta}^-(E) - f_\alpha(E)[1 - f_\beta(E + \omega)] T_{\alpha,\beta}^+(E) \mathrm{d}E \tag{5}$$

$$T_{\alpha,\beta}^-(E) = N\mathrm{Tr}\{\Gamma_\alpha(E)G(E)MA_\beta(E - \omega)M^\dagger G^\dagger(E)\} \tag{6}$$

$$T_{\alpha,\beta}^+(E) = N\mathrm{Tr}\{\Gamma_\alpha(E)G(E)M^\dagger A_\beta(E + \omega)MG^\dagger(E)\} \tag{7}$$

where $f_\alpha$ is the Fermi Dirac distribution function of electrode α, *G* and $G^\dagger$ are the retarded and advanced Green's functions, and *M* the electron−photon coupling matrix given by

$$M_{ml} = \frac{e}{m_0} \left( \frac{\sqrt{\mu_r \epsilon_r}}{2N\omega\epsilon c} F \right)^{1/2} \mathbf{e} \cdot \mathbf{P}_{ml} \tag{8}$$

where $\epsilon, \epsilon_r,$ and $\mu_r$ refer to the isotropic homogeneous permittivity, relative permittivity, and relative permeability, respectively, *F* is the photo flux, **e** is a unit vector giving the polarization of the light, and $\mathbf{P}_{ml}$ is the momentum operator. The total photocurrent is then given by $I_{ph} = I_D - I_S$. In the present work, the photon energy from 0 to 5 eV is detected and the linearly polarized light along the transport direction is considered.

Figure 6(c) shows the maps of photocurrent density as a function of photon energy for the z-$B_4H_4$ and a-$B_4H_4$ under zero drain-source bias. Interestingly, upon illumination, $B_4H_4$ can generate observable photocurrents along both the zigzag and armchair directions. The total photoexcited current density per second is about 26.1 $\mu A/cm^2$ along the zigzag direction under the AM1.5 standard [46], significantly larger than along the armchair one (about 0.4 $\mu A/cm^2$), which can be measured experimentally with a proper laser power [54,55]. $B_4H_4$ ML shows high anisotropy in the photocurrents along the zigzag and armchair directions, analogous to the phosphorene photodetector [48]. Differently, phosphorene photodetector has larger photocurrent along the armchair direction, which is up to two orders of magnitude higher than that along the zigzag one. Under the visible light region, both the z-$B_4H_4$ and a-$B_4H_4$ have a photocurrent peak, and they are excited at blue (green) light for the z-$B_4H_4$ (a-$B_4H_4$). This is mainly because that both the z-$B_4H_4$ and a-$B_4H_4$ have large absorption coefficient to that light with the corresponding color. Thus, $B_4H_4$ can be used as a photodetector based on its different photoelectronic response along the zigzag and armchair directions. Recent success in fabrication of photodetectors based on 2D materials, including $PdSe_2$ [49], graphene-based heterojunctions [45,50], and black phosphorous [51], indicates that the use of $B_4H_4$-based as photodetector is also possible. To detect the photoelectronic response along the two directions (i.e., zigzag and armchair), one may obtain strong signals under the illumination of blue and green lights, respectively.

**IV. CONCLUSION**

In conclusion, by means of the first-principles calculations, we systematically investigate the electronic transport and photoelectronic properties of 2D $B_4H_4$ ML along the zigzag and armchair directions. Our results show that $B_4H_4$ ML presents a promising current-limiting effect and possess two optional levels along the zigzag direction (high-level) and armchair one (low-level) because of its strong electrical anisotropy. Furthermore, $B_4H_4$ can generate sizable photocurrents under illumination, with strong photon absorption to blue/green light along the zigzag/armchair direction. Our results unveil that the $B_4H_4$ ML is promising for the utilization in current limiters and photodetectors.

ACKNOWLEDGMENTS

The work at Henan Normal University was supported by the National Natural Science Foundation of China (Nos. 11774079, U1704136, 11874429 and 11774179), the CSC (No. 201708410368), the Natural Science Foundation of Henan Province (No. 162300410171), the young backbone teacher training program of Henan province's higher education (No. 2017GGJS043), the Science Foundation for the Excellent Youth Scholars of Henan Normal University (No. 2016YQ05), the Henan Overseas Expertise Introduction Center for Discipline Innovation (No. CXJD2019005), and the Project of High-level Talents of Hunan Province (No. 2018RS3021). The work at the University of California at Irvine was supported by the US DOE-BES under Grant DE-FG02-05ER46237. We also thank X. Dai at Zhengzhou Normal University for helpful discussion, and the High-Performance Computing Centre of Henan Normal University.

REFERENCES

[1] A. J. Mannix, X. F. Zhou, B. Kiraly, J. D. Wood, D. Alduicn, B. D. Myers, X. Liu, B. L. Fisher, U. Santiago, J. R. Guest *et al.*, Synthesis of borophenes: Anisotropic,two-dimensional boron polymorphs, Science **350**, 1513 (2015).

[2] B. Feng, J. Zhang, Q. Zhong, W. Li, S. Li, H. Li, C. Peng, S. Meng, L. Chen, K. Wu, Experimental realization of two-dimensional boron sheets, Nature Chem. **8**, 563 (2016).

[3] L. Kou, Y. Ma, C. Tang, Z. Sun, A. Du, and C. Chen, Auxetic and ferroelastic borophane: A novel 2D material with negative Possion's ratio and switchable dirac transport channels, Nano Lett. **16**, 7910−7914 (2016).

[4] J. Li, Y. Wei, X. Fan, H. Wang, Y. Song, G. Chen, Y. Liang, V. Wang, and Y. Kawazoe, Global minimum of two-dimensional $FeB_6$ and an oxidization induced negative Poisson's ratio: A new stable allotrope, J. Mater. Chem. C **4**, 9613 (2016).

[5] A. Lopez-Bezanilla, Twelve inequivalent Dirac cones in two-dimensional $ZrB_2$, Phys.

Rev. Mater. **2**, 011002 (2018).

[6] J. C. Alvarez-Quiceno, R. H. Miwa, G. M. Dalpian, and A. Fazzio, Oxidation of free-standing and supported borophene, 2D Mater. **4**, 025025 (2017).

[7] Z.-Q. Wang, T.-Y. Lü, H.-Q. Wang, Y. P. Feng, and J.-C. Zheng, Review of borophene and its potential applications, Front. Phys. **14**, 33403 (2019).

[8] J.-H. Yang, S. Song, S. Du, H.-J. Gao, and B. I. Yakobson, Design of two-dimensional graphene-like Dirac materials $\beta_{12}$-XBeB$_5$ (X = H, F, Cl) from non-graphene-like $\beta_{12}$-borophene, J. Phys. Chem. Lett. **8**, 4594−4599 (2017).

[9] L. Zhang, P. Liang, H.-b. Shu, X.-l. Man, F. Li, J. Huang, Q.-m. Dong, and D.-l. Chao, Borophene as efficient sulfur hosts for lithium–sulfur batteries: Suppressing shuttle effect and improving conductivity, J. Phys. Chem. C **121**, 15549−15555 (2017).

[10] J. Li, X. Fan, Y. Wei, J. Liu, J. Guo, X. Li, V. Wang, Y. Liang, and G. Chen, Voltage-gated spin-filtering properties and global minimum of planar MnB$_6$, and half-metallicity and room-temperature ferromagnetism of its oxide sheet, J. Mater. Chem. C **4**, 10866−10875 (2016).

[11] Y. Zhao, S. Zeng, and J. Ni, Phonon-mediated superconductivity in borophenes, Appl. Phys. Lett. **108**, 242601 (2016).

[12] L. Yuan, L. Kang, Y. Chen, D. Wang, J. Gong, C. Wang, M. Zhang, and X. Wu, Hydrogen storage capacity on Ti-decorated porous graphene: First-principles investigation, Appl. Surf. Sci. **434**, 843−849 (2018).

[13] S. H. Mir, S. Chakraborty, P. C. Jha, J. Wärnå, H. Soni, P. K. Jha, and R. Ahuja, Two-dimensional boron: Lightest catalyst for hydrogen and oxygen evolution reaction, Appl. Phys. Lett. **109**, 053903 (2016).

[14] L.-C. Xu, A. Du, and L. Kou, Hydrogenated borophene as a stable two-dimensional Dirac material with an ultrahigh Fermi velocity, Phys. Chem. Chem. Phys. **18**, 27284 (2016).

[15] N. K. Jena, R. B. Araujo, V. Shukla, and R. Ahuja, Borophane as a benchmate of graphene: A potential 2D material for anode of Li and Na-Ion batteries, ACS Appl. Mater. Interfaces **9**, 16148 (2017).

[16] H. Zhang, Y. Li, J. Hou, A. Du, and Z. Chen, Dirac state in the FeB$_2$ monolayer with graphene-like boron sheet, Nano Lett. **16**, 6124 (2016).

[17] A. Lopez-Bezanilla, Interplay between p− and d− orbitals yields multiple Dirac states in one-and two-dimensional CrB$_4$, 2D Mater. **5**, 035041 (2018).

[18] X. Qu, J. Yang, Y. Wang, J. Lv, Z. Chen, and Y. Ma, A two-dimensional TiB$_4$ monolayer exhibits planar octacoordinate Ti, Nanoscale **9**, 17983−17990 (2017).

[19] M. Martinez-Canales, T. R. Galeev, A. I. Boldyrev, and C. J. Pickard, Dirac cones in two-dimensional borane, Phys. Rev. B **96**, 195442 (2017).

[20] Y. Jiao, F. Ma, J. Bell, A. Bilic, and A. Du, Two-dimensional boron hydride sheets: High stability, massless Dirac Fermions, and excellent mechanical properties, Angew. Chem. **128**, 10448−10451 (2016).

[21] B. Mortazavi, M. Makaremi, M. Shahrokhi, M. Raeisi, C. V. Singh, T. Rabczuk, and L. F. C. Pereira, Borophene hydride: a stiff 2D material with high thermal conductivity and attractive optical and electronic properties, Nanoscale **10**, 3759−3768 (2018).

[22] H. Nishino, T. Fujita, N. T. Cuong, S. Tominaka, M. Miyauchi, S. Iimura, A. Hirata, N. Umezawa, S. Okada, E. Nishibori, A. Fujino, T. Fujimori, S. Ito, J. Nakamura, H. Hosono, T. Kondo, Formation and characterization of hydrogen boride sheets derived from MgB$_2$ by cation exchange, J. Am. Chem. Soc. **139**, 13761 (2017).

[23] J. Taylor, H. Guo, and J. Wang, Ab initio modeling of open systems: Charge transfer, electron conduction, and molecular switching of a C$_{60}$ device, Phys. Rev. B **63**, 121104(R) (2001).

[24] M. Brandbyge, J.-L. Mozos, P. Ordejón, J. Taylor, and K. Stokbro, Density-functional method for nonequilibrium electron transport, Phys. Rev. B **65**, 165401 (2002).

[25] J. M. Soler, E. Artacho, J. D. Gale, A. García, J. Junquera, P. Ordejón, and D. Sánchez-Portal, The SIESTA method for *ab initio* order-N materials simulation, J. Phys.: Condens. Matter. **14**, 2745 (2002).

[26] J. P. Perdew, J. A. Chevary, S. H. Vosko, K. A. Jackson, M. R. Pederson, D. J. Singh, and C. Fiolhais, Atoms, molecules, solids, and surfaces: Applications of the generalized gradient approximation for exchange and correlation, Phys. Rev. B **46**, 6671 (1992).

[27] J. P. Perdew, K. Burke, and M. Ernzerhof, Generalized gradient approximation made simple, Phys. Rev. Lett. **77**, 3865 (1996).

[28] M. Schlipf and F. Gygi, Optimization algorithm for the generation of ONCV

pseudopotentials, Comput. Phys. Commun. **196**, 36 (2015).

[29] L. Zhang, K. Gong, J. Chen, L. Liu, Y. Zhu, D. Xiao, and H. Guo, Generation and transport of valley-polarized current in transition-metal dichalcogenides, Phys. Rev. B **90**, 195428 (2014).

[30] Y. Xie, M. Chen, Z. Wu, Y. Hu, Y. Wang, J. Wang, and H. Guo, Two-dimensional photogalvanic spin-battery, Phys. Rev. Appl. **10**, 034005 (2018).

[31] R. Quhe, Q. Li, Q. Zhang, Y. Wang, H. Zhang, D. Chen, K. Liu, Y. Yu, D. Lun, P. Feng *et al.*, Simulations of quantum transport in Sub-5-nm monolayer phosphorene transistors, Phys. Rev. Appl. **10**, 024022 (2018).

[32] M. Büttiker, Y. Imry, R. Landauer, and S. Pinhas, Generalized many-channel conductance formula with application to small rings, Phys. Rev. B **31**, 6207 (1985).

[33] Z. Li, H. Qian, J. Wu, B.-L. Gu, and W. Duan, Role of symmetry in the transport properties of graphene nanoribbons under bias, Phys. Rev. Lett. **100**, 206802 (2008).

[34] Y. P. An, M. J. Zhang, D. P. Wu, Z. M. Fu, and K. Wang, The electronic transport properties of transition-metal dichalcogenide lateral heterojunctions, J. Mater. Chem. C **4**, 10962 (2016).

[35] Y. P. An, M. J. Zhang, H. X. Da, Z. M. Fu, Z. Y. Jiao, and Z. Y. Liu, Width and defect effects on the electronic transport of zigzag $MoS_2$ nanoribbons, J. Phys. D: Appl. Phys. **49**, 245304 (2016).

[36] V. Shukla, A. Grigoriev, N. K. Jena, and R. Ahuja, Strain controlled electronic and transport anisotropies in two-dimensional borophene sheets, Phys. Chem. Chem. Phys. **20**, 22952 (2018).

[37] S. Datta, Electronic Transport in Mesoscopic Systems, (Cambridge University Press, Cambridge, England, 1995).

[38] Y. An, J. Jiao, Y. Hou, H. Wang, R. Wu, C. Liu, X. Chen, T. Wang, and K. Wang, Negative differential conductance effect and electrical anisotropy of 2D $ZrB_2$ monolayers, J. Phys.: Condens. Matt. **31**, 065301 (2019).

[39] G. C. Solomon, C. Herrmann, T. Hansen, V. Mujica, and M. A. Ratner, Exploring local currents in molecular junctions, Nature Chem. **2**, 223 (2010).

[40] A. H. Castro Neto, F. Guinea, N. M. R. Peres, K. S. Novoselov, and A. K. Geim, The

electronic properties of graphene, Rev. Mod. Phys. **81**, 109 (2009).

[41] S. Zamani and R. Farghadan, Graphene nanoribbon spin-photodetector, Phys. Rev. Appl. **10**, 034059 (2018).

[42] C.-C. Tang, K. Ikushima, D. C. Ling, C. C. Chi, and J.-C. Chen, Quantum hall dual-band infrared photodetector, Phys. Rev. Appl. **8**, 064001 (2017).

[43] F. Xia, T. Mueller, Y.-m. Lin, A. Valdes-Garcia, and P. Avouris, Ultrafast graphene photodetector, Nature Nanotech. **4**, 839−843 (2009).

[44] J. Li, C. Zhao, B. Liu, C. You, F. Chu, N. Tian, Y. Chen, S. Li, B. An, A. Cui, Metamaterial grating-integrated graphene photodetector with broadband high responsivity, Applied Surface Science **473**, 633 (2019).

[45] Q. Zhou, J. Shen, X. Liu, Z. Li, H. Jiang, S. Feng, W. Feng, Y. Wang, and D. Wei, Hybrid graphene heterojunction photodetector with high infrared responsivity through barrier tailoring, Nanotech. **30**, 195202 (2019).

[46] ASTM, Astm **03**, 1–21 (2013).

[47] M. Palsgaard, T. Gunst, T. Markussen, K. S. Thygesen, and M. Brandbyge, Stacked janus device concepts: Abrupt pn-junctions and cross-plane channels, Nano Lett. **18**, 7275−7281 (2018).

[48] S. Li, T. Wang, X. Chen, W. Lu, Y. Xie, and Y. Hu, Self-powered photogalvanic phosphorene photodetectors with high polarization sensitivity and suppressed dark current, Nanoscale **10**, 7694−7701 (2018).

[49] L.-H. Zeng, D. Wu, S. H. Lin, C. Xie, H. Y. Yuan, W. Lu, S. P. Lau, Y. Chai, L. B. Luo, Z. J. Li, Controlled synthesis of 2D palladium diselenide for sensitive photodetector applications, Adv. Funct. Mater. **29**, 1806878 (2019).

[50] T. Yu, F. Wang, Y. Xu, L. Ma, X. Pi, and D. Yang, Graphene coupled with silicon quantum dots for high-performance bulk-silicon-based schottky-junction photodetectors, Adv. Mater. **28**, 4912 (2016).

[51] Q. Guo, A. Pospischil, M. Bhuiyan, H. Jiang, H. Tian, D. Farmer, B. Deng, C. Li, S. Han, H. Wang, Black phosphorus mid-infrared photodetectors with high gain, Nano Lett. **16**, 4648 (2016).

[52] L. E. Henrickson, Nonequilibrium photocurrent modeling in resonant tunneling

photodetectors, J. Appl. Phys. **91**, 6273−6281 (2002).

[53] J. Chen, Y. Hu, and H. Guo, First-principles analysis of photocurrent in graphene PN junctions, Phys. Rev. B **85**, 155441 (2012).

[54] J. Wu, G. K. W. Koon, D. Xiang, H. Han, C. T. Toh, E. S. Kulkarni, I. Verzhbitskiy, A. Carvalho, A. Rodin, S. P. Koenig, Colossal ultraviolet photoresponsivity of few-Layer black phosphorus, ACS Nano **9**, 8070−8077 (2015).

[55] T. Hong, B. Chamlagain, W. Lin, H.-J. Chuang, M. Pan, Z. Zhou, and Y.-Q. Xu, Polarized photocurrent response in black phosphorus field-effect transistors, Nanoscale **6**, 8978−8983 (2014).